\documentclass[a4paper]{article}
\usepackage[pdftex,
     colorlinks=false]{hyperref}

\usepackage[fleqn]{amsmath}
\usepackage{amssymb}
\usepackage{bm}
\usepackage{mathrsfs}
\usepackage{graphicx}
\usepackage{color}

\usepackage[]{espcrc2}

\numberwithin{equation}{section}

\DeclareMathOperator{\id}{id}

\DeclareMathOperator{\SOrth}{SO}
\DeclareMathOperator{\SL}{SL}

\DeclareMathOperator{\Orthframe}{\mathcal{F}}
\DeclareMathOperator{\Redframe}{\Orthframe_{\text{red}}}
\DeclareMathOperator{\Lieder}{\mathscr{L}}

\newcommand{\bnabla}{\bm{\nabla}}
\newcommand{\orbit}{\mathcal{O}}

\newcommand{\bbR}{\mathbb{R}}

\newcommand{\half}{\frac{1}{2}}

\newcommand{\vece}{\bm{e}}
\newcommand{\vecE}{\bm{E}}

\newcommand{\bbZ}{\mathbb{Z}}
\DeclareMathOperator{\adS}{adS}
\DeclareMathOperator{\dS}{dS}

\title{Black Holes Without Coordinates}

\author{Orlando Alvarez%
    \address{Department of Physics\\
    University of Miami\\
    P.O. Box 248046\\
    Coral Gables, FL 33146 USA}%
    \thanks{email: \texttt{oalvarez@miami.edu}. This work was supported in
    part by the National Science Foundation under Grants PHY-0244261
    and PHY-0554821.}
}

\begin{document}

\begin{abstract}
    These lectures describe how to study the geometry of some 
    black holes without the use of coordinates.
    
    Lectures presented at the Carges\`{e} Summer School, ``Theory and
    Particle Physics: the LHC perspective and beyond'' (June 16 to
    June 28, 2008). This paper is a sequel to \texttt{gr-qc/0701115}.
\end{abstract}

\maketitle

\section{Introduction}
\label{sec:intro}

In these lectures I discuss how to study some solutions to the
Einstein equations in a coordinate independent manner.  The main ideas
were presented lecture style in reference \cite{Alvarez:2007xs} where
the Schwarzschild solution was studied in detail.  Here I will develop
some background material on the frame bundles that was implicit in
\cite{Alvarez:2007xs} and present some unpublished studies about the
black hole discovered by Ba\~{n}ados, Teitelboim and Zanelli (BTZ)
\cite{Banados:1992wn} that was of great interest at this
Institute.  These lectures should be viewed as an addendum to the
written up Schwarzschild discussion.

\section{Frame Bundles}
\label{sec:frame}

In general there is no global coordinate system or global frame you
can impose on a manifold.  If a manifold admits a global framing then
it is called parallelizable and the list of such manifolds is small.
For us the important fact is that every manifold $M$ has an associated
parallelizable manifold called the bundle of frames.  E.~Cartan showed
how to reconstruct the geometry of the manifold by studying the geometry of
the frame bundle. We are are interested in semi-riemannian geometry 
where there is a metric on the manifold. The discussion that follows 
is the same in the strictly riemannian or the lorentzian case and we 
give it using the language of the former.

Assume we have an $n$-dimensional manifold $M$ with metric,
\emph{i.e.}, for any pair of tangent vectors on $M$ we know how take
their inner product.  We can always locally find an orthonormal frame
of tangent vector fields but in general this cannot be extended
globally.  We can construct a fiber bundle whose local section are
orthonormal frames.  This bundle is called the orthonormal frame
bundle $\Orthframe(M)$ of the manifold.  It is a remarkable
mathematical fact that this bundle is parallelizable, \emph{i.e.}, it
admits a global framing.  We discuss the construction of this bundle.

Let $\{U_{\alpha}\}$ be an open cover of $M$ such that on each
$U_{\alpha}$ we can choose a fiducial orthonormal frame $E_{\alpha} =
(\vecE_{1},\vecE_{2},\ldots,\vecE_{n})_{\alpha}$ that we write as a
row vector.  On the overlap $U_{\alpha} \cap U_{\beta}$, the fiducial
frames are related by $E_{\alpha} = E_{\beta} \varphi_{\beta\alpha}$
where $\varphi_{\beta\alpha} : U_{\alpha} \cap U_{\beta} \to
\SOrth(n)$.  This is just the statement that we have a metric.  For
simplicity we assume our manifold is orientable (and also time
orientable in the lorentzian case) and this means that the
``transition functions'' $\varphi_{\beta\alpha}$ may be restricted to
orthogonal matrices with determinant one (and also preserve the time
orientation).  We have that $\varphi_{\alpha\beta} =
\varphi_{\beta\alpha}^{-1}$ and $\varphi_{\alpha\beta}
\varphi_{\beta\gamma} \varphi_{\gamma\alpha}=I$.  If $e_{\alpha}$ is
another orthonormal frame at $x\in U_{\alpha}$ then there exists a
unique orthogonal matrix $g_{\alpha}$ such that $e_{\alpha} =
E_{\alpha} g_{\alpha}$.  Therefore the set of all orthonormal frames
over $U_{\alpha}$ is isomorphic to $U_{\alpha}\times \SOrth(n)$.  The
idea is to put all the $U_{\alpha}\times \SOrth(n)$ together to make a
bundle.  To do this we require that if $x \in U_{\alpha} \cap
U_{\beta}$ then we identify $(x,g_{\alpha})$ with $(x,g_{\beta})$ via
$g_{\alpha} = \varphi_{\alpha\beta}(x) g_{\beta}$.  This constructs a
bundle $\pi: \Orthframe(M) \to M$ with fiber isomorphic to $\SOrth(n)$
called the bundle of orthonormal frames.  We have that $\dim
\Orthframe(M) = n + \tfrac{1}{2}n(n-1) = \tfrac{1}{2}n(n+1)$.

We now show this bundle has a global framing.  It is simpler to
construct a coframing and we do this.  On $U_{\alpha} \subset M$, let
$\vartheta_{\alpha}$ be the frame dual to $E_{\alpha}$.  The coframe
$\vartheta_{\alpha}$ is taken to be a row vector of $1$-forms on $M$.
The Levi-Civita connection $\varpi_{\alpha}$ is an anti-symmetric
matrix of $1$-forms on $U_{\alpha}$ that satisfy $d\vartheta_{\alpha} 
= -\varpi_{\alpha} \wedge \vartheta_{\alpha}$. On $U_{\alpha} \times 
\SOrth(n)$ define $1$-forms by using the pullback $\pi^{*}$:
\begin{equation}
    \begin{split}
    \theta_{\alpha} &= g_{\alpha}^{-1}\,\pi^{*}\vartheta_{\alpha}\,, \\
    \omega_{\alpha} &= g_{\alpha}^{-1}\, dg_{\alpha} + g_{\alpha}^{-1} 
    (\pi^{*} \varpi_{\alpha})g_{\alpha}\,.
    \end{split}
    \label{eq:def-theta-omega}
\end{equation}
Note that the matrix of forms $\omega_{\alpha}$ is antisymmetric.
This is a coframing of $U_{\alpha} \times \SOrth(n)$ because the
$\theta_{\alpha}$ tells you about ``horizontal'' motion and
$g_{\alpha}^{-1}\, dg_{\alpha}$ measures vertical motion.  The
$\theta_{\alpha}$ and the $\omega_{\alpha}$ are linearly independent
$1$-forms on $U_{\alpha} \times \SOrth(n)$.  If $\sigma: U_{\alpha}
\to U_{\alpha} \times \SOrth(n)$ is a local section then $\sigma: x
\mapsto (x,g(x))$.  This section gives a local framing $e_{\alpha}(x)
= E_{\alpha}(x) g(x)$ with dual framing $g(x)^{-1} \vartheta(x)$ and
Levi-Civita connection $g^{-1}dg + g^{-1}\varpi g$.  Using
\eqref{eq:def-theta-omega} we see that we can obtain the same results
by noting that the dual frame is $\sigma^{*}\theta_{\alpha}$ and the
local Levi-Civita connection is $\sigma^{*}\omega_{\alpha}$.  To show
this we need that a section satisfies $\pi\circ\sigma =\id$.  Note
that the $1$-forms $\sigma^{*}\theta_{\alpha}$ and
$\sigma^{*}\omega_{\alpha}$ are linearly dependent $1$-forms on
$U_{\alpha}$.

It is not clear why this is useful until you observe that on overlaps
$U_{\alpha}\cap U_{\beta}$ we have $\theta_{\alpha} = \theta_{\beta}$ 
and $\omega_{\alpha} = \omega_{\beta}$ and therefore these define 
global $1$-forms $\theta$, $\omega$ on $\Orthframe(M)$. This 
is the global coframing of the frame bundle. This was discovered by 
E.~Cartan. He pointed out that this coframe satisfies the structural 
equations
\begin{align}
    d\theta^{\mu} & = - \omega^{\mu}{}_{\nu} \wedge \theta^{\nu}\,,
    \label{eq:C-1}  \\
    d\omega^{\mu}{}_{\nu} & = -\omega^{\mu}{}_{\lambda} \wedge 
    \omega^{\lambda}{}_{\nu} + \tfrac{1}{2}\, 
    R^{\mu}{}_{\nu\kappa\lambda} \theta^{\kappa} \wedge 
    \theta^{\lambda}\,.
    \label{eq:C-2}
\end{align}
The curvature functions $R_{\mu\nu\kappa\lambda}$ are ordinary
functions on the frame bundle that transforms nicely as you move up
and down a fiber, \emph{i.e.}, under the action of $\SOrth(n)$.  These
are globally defined functions on the frame bundle.  If we consider a
local section $s$ of the frame bundle then the pullback functions
$s^{*}R_{\mu\nu\kappa\lambda}$ on $U_{\alpha}$ are the components of
the curvature tensor with respect to the coframe $s^{*}\theta$.

The main idea is to use the global coframing on the frame bundle to 
study the global geometry of black holes. In this way we will avoid 
the introduction of coordinate singularities or other confusions.

At this point the reader can read \cite{Alvarez:2007xs} and see how
these ideas can be applied to the study of the Schwarzschild
spacetime.  He we will adapt the ideas presented and the notation used
in that paper to study the BTZ black hole.

\section{Einstein Equations with a Cosmological Constant}
\label{sec:vac-E-eqs}

The BTZ black hole is a black hole solution in $(1+2)$-dimensional
gravity on a lorentzian manifold $N$ with a cosmological constant.
The vacuum Einstein equations with cosmological constant are
\begin{equation}
    R^{N}_{\mu\nu}- \frac{1}{2}\,R^{N}\eta_{\mu\nu} + \Lambda 
    \eta_{\mu\nu} =0\,.
    \label{eq:E-eqs}
\end{equation}
Taking the trace we see that $-R^{N}/2 + 3\Lambda=0$ and therefore we 
can write the above as
\begin{equation}
    R^{N}_{\mu\nu} -2\Lambda \eta_{\mu\nu}=0\,.
    \label{eq:E-eqs-1}
\end{equation}
Since $\dim N=3$, the Ricci tensor determines the full curvature
tensor and therefore we see that we have a constant curvature manifold
with
\begin{equation}
    R^{N}_{\mu\nu\rho\sigma} = \Lambda \left( 
    \eta_{\mu\rho}\eta_{\nu\sigma} - \eta_{\mu\sigma}\eta_{\nu\rho} 
    \right).
    \label{eq:N-cons-curv}
\end{equation}
This is a general result special to $\dim N=3$. This is the source of 
the big conundrum posed by the BTZ black hole. Since the curvature is 
fully determined there is no room for gravitational waves (degrees of 
freedom) but the BTZ black hole has a temperature and thus an 
entropy. There should be degrees of freedom. What is going on? This 
was discussed by others in this Institute.

We conclude with the following observations about the local geometry
of $N$.
\begin{enumerate}
    
    \item ($\Lambda=0$) $N$  is locally isometric to $\mathbb{M}^{3}$.

    \item ($\Lambda>0$)  $N$ is locally isometric to deSitter space $\dS_{3}$.

    \item ($\Lambda<0$) $N$ is locally isometric to anti-deSitter
    space $\adS_{3}$.
\end{enumerate}
Let $\widetilde{N}$ be the simply connected universal cover of $N$.
This is a simply connected maximally symmetric manifold with a
transitive group of isometries and it looks the same everywhere.  This
cannot be a black hole because of the homogeneity.  If $D$ is a group
of discrete isometries acting on $\widetilde{N}$ then
$N=\widetilde{N}/D$ may be an interesting Lorentz manifold.  In
general this manifold will not admit a transitive group of isometries
and thus different parts of the manifold will have distinct
properties.  In fact, this is how the BTZ solution arises
\cite{Banados:1992gq}.  The geometry of the BTZ solution is discussed
in great detail in this article. 

We are very familiar with an analogous example.  Riemann surfaces with
genus greater than one are the quotient of the upper half plane by a
discrete subgroup of $\SL(2,\bbR)$.  The upper half plane with the
Poincar\'{e} metric admits $\SL(2,\bbR)$ as a transitive group of
isometries.  After identifying points using the discrete subgroup we
get a torus with at least two holes.  Such a manifold does not admit
Killing vectors and may only have discrete isometries.  It is very
different from the upper half plane even though locally they look 
alike  because they have the same local curvature.

A brief description of  $\adS$ manifolds is given in 
Appendix~\ref{sec:adS3-Basics}.

\section{Circularly Symmetric $(1+2)$ Geometry}
\label{sec:birkhoff}

Assume $N$ is a $3$-dimensional lorentzian manifold that is both
orientable and time orientable.  This means that the structure group
of of the orthonormal Lorentz frame bundle is
$\SOrth^{\uparrow}(1,2)$, the connected component of the Lorentz
group.  We assume the spacetime is a semi-riemannian submersion $\pi:
N \to M$ with the fiber being a space-like $1$-dimensional manifold.
The existence of the vertical distribution of vector fields means that
the structure group of the reduced frame bundle $\Redframe(N)$ is
reduced to $\SOrth^{\uparrow}(1,1)$.  A consequence is that there are
no $\omega_{ab}$ and only one $\pi_{ij}$.  If we use some type of
``Schwarzschild spherical coordinates'' denoted by $(t,r,\phi)$.  Then
we will only have non-vanishing connection $\pi_{tr}$.

We have a pseudo-riemannian submersion.  We denote $\theta^{2}$ by 
$\varphi$. The full structural equations
for a riemannian submersion are
\begin{equation}
    \begin{split}
	K_{22i} &= K_{i}\,, \quad A_{ij} =A \epsilon_{ij}\,, \\
	\omega_{2i} &= K_{i}\varphi - A_{ij}\theta^{j}\,,  \\
	\pi_{ij} &= \omega_{ij} -A_{ij}\varphi\,, \\
	d\theta^{i} &= -\pi_{ij}\wedge\theta^{j}\,,  \\
	d\varphi &= 
	-K_{i}\varphi\wedge\theta^{i} - 
	2A \theta^{0}\wedge\theta^{1}\,,\\
	d\pi_{ij} &= -\pi_{ik}\wedge\pi_{kj} + \half R^{M}{}_{ijkl}\, 
	\theta^{k}\wedge\theta^{l}\,, \\
    \end{split}
    \label{eq:submersion-full}
\end{equation}
Being more explicit we have
\begin{align}
    d\theta^{0} & = +\pi \wedge \theta^{1}\,,
    \label{eq:SO2-dtheta0}  \\
    d\theta^{1} & = +\pi \wedge \theta^{0}\,,
    \label{eq:SO2-dtheta1} \\
    d\pi & = k^{M}\; \theta^{0} \wedge \theta^{1}\,,
    \quad\text{where}\quad \pi= \pi_{01}\,,
    \label{eq:SO2-dpi} \\
    d\varphi & = -K_{i}\varphi\wedge\theta^{i} - 
	2A \theta^{0}\wedge\theta^{1}\,.
    \label{eq:SO2-dvarphi}
\end{align}
We note that $dA = A_{;i}\theta^{i} + A_{;\varphi}\phi$ and $dK_{i} = 
-\pi_{ij}K^{j} + K_{i;j}\theta^{j} + K_{i;\varphi}\varphi$. From 
$d^{2}\pi=0$ we learn that $dk^{M}= k^{M}_{;j}\theta^{j}$, 
\emph{i.e.}, $k^{M}$ is the pullback of a function on $M$. From 
$d^{2}\varphi=0$ we learn that
\begin{equation*}
    0 = \left( -K_{0;1} + K_{1;0} -2 A_{;\varphi}\right) 
    \theta^{0}\wedge \theta^{1}\wedge \varphi\,.
\end{equation*}
This tells us that
\begin{equation}
    K_{0;1} - K_{1;0} = -2 A_{;\varphi}\,.
    \label{eq:K-curl}
\end{equation}
In other words $d(K_{i}\theta^{i}) = 
-2A_{;\varphi}\, \theta^{0}\wedge\theta^{1}$.

Next we look at the Ricci tensor:
\begin{align}
    R^{N}_{\varphi\varphi} & = -K^{i}{}_{;i} -K_{i}K^{i} - 2A^{2}\,,
    \label{eq:R-ff}  \\
    R^{N}_{\varphi i} & = -\epsilon_{i}{}^{j} A_{;j} - 2 
    \epsilon_{ij}K^{j}A\,,
    \label{eq:R-fi}  \\
    R^{N}_{ij} & = -k^{M}\eta_{ij} - A^{2}\eta_{ij} -K_{i}K_{j} - 
    \frac{1}{2} \left( K_{i;j} + K_{j;i}\right).
    \label{eq:R-ij}
\end{align}

Next, we assume there is an $\SOrth(2)$ action that leaves the
metric invariant and that the orbit of a point is a $1$-dimensional
spacelike circle.  Let $\orbit_{p}$ be the orbit through $p \in N$.
This action leads to a foliation of $N$ by the $1$-dimensional orbits.
Under some assumptions of a constant dimensionality of the orbits we
can assume that this foliation is actually a fibration.  Our
hypothesis tells us that $\dim\orbit_{p} = 2$.  If $G_{p}$ is the
isotropy group at $p$ then $\dim G_{p} = 0$, \emph{i.e.}, $G_{p}
\approx \bbZ_{l}$.  This tells us that $\orbit_{p} \approx
\SOrth(2)/G_{p} \approx S^{1}$.      If $\pi:N\to M$ is
our fiber bundle and if $\pi(p) = x$ then the fiber over $x$ is given
by $F_{x} = \orbit_{p}$.

At $p \in N$ we can write
$T_{p}N = T_{p}\orbit_{p} \oplus T_{p}\orbit_{p}^{\perp}$ and the
$\SOrth(2)$ action tells us that both the riemannian metric on
$T_{p}\orbit_{p}$ and the lorentzian metric on
$T_{p}\orbit_{p}^{\perp}$ are invariant under the $\SOrth(2)$ action.
At $p \in N$, all geometrical structures must be invariant under the
isotropy group action $G_{p} \approx \bbZ_{l}$.  The action of $G_{p}$
on $T_{p} \orbit_{p}$ is trivial.  To see this let parametrize the
points of $\orbit_{p}$ as $e^{i\phi}$ then the action of $\SOrth(2)$
is of the form $e^{i\phi} \mapsto e^{i\phi'} = e^{ir\theta}e^{i\phi}$.
From this we see that $d\phi'=d\phi$ and this implies that the
isotropy group action is trivial on $T_{p}\orbit_{p}$.  The action on
$T_{p}\orbit_{p}^{\perp}$ is automatically trivial because there is no
$\bbZ_{l}$ subgroup in $\SOrth^{\uparrow}(1,1)$.

Next we explore additional properties that follow from the $\SOrth(2)$
action.  First we observe that $\omega_{ab}$ did not get modified by
the symmetry breakdown and therefore the
$\SOrth(2)$ Killing vector has the form
\begin{equation}
    V = V^{\varphi} \vece_{\varphi}\,.
    \label{eq:Killing-V-exp}
\end{equation}
because ``$\omega_{\varphi\varphi}=0$'', see the discussion around
equation~(6.2) in reference \cite{Alvarez:2007xs}.  From
$\Lieder_{V}\varphi=0$ learn that
\begin{equation}
    dV^{\varphi} = V^{\varphi}\, K_{i}\theta^{i}\,.
    \label{eq:dVf}
\end{equation}
Using $0 = d\Lieder_{V}\varphi = \Lieder_{V}(d\varphi)$ leads to
\begin{equation}
    V(K_{i})=0\,,\quad\text{and}\quad V(A)=0.
    \label{eq:constancy-fiber}
\end{equation}
The functions $K_{i}$ and $A$ on $\Redframe(N)$ are constant along each
orbit $\orbit_{p}$.  This means that
\begin{equation}
    dA = A_{;i}\theta^{i}\,,\quad\text{and}\quad dK_{i} = 
    -\pi_{ij}K^{j} + K_{i;j}\theta^{j}\,.
    \label{eq:der-A-K}
\end{equation}
In other words we have $A_{;\varphi}=0$ and $K_{i;\varphi}=0$. Going 
back to \eqref{eq:K-curl} we see that the group action tells us that
\begin{equation}
    d(K_{i}\theta^{i})=0\,.
    \label{eq:K-curl-0}
\end{equation}
This means that locally we can find a function $f$ on $\Redframe(N)$
such that $df = K_{i}\theta^{i}$.  In fact we can do better that this.
Choose $p \in N$ and lets look at a small tubular neighborhood of
$\orbit_{p}$.  At $p$ choose a small disk transverse to $\orbit_{p}$
with local coordinates $(y^{0},y^{1})$.  The third coordinate is
generated by the action of $\SOrth(2)$ at $(y^{0},y^{1})$ and in this
way we coordinatize the tubular neighborhood locally by coordinates
$(y^{0},y^{1},\phi)$ where $\phi \in [0,2\pi)$ is the standard
coordinate on a circle.  The local submersion geometry tells us that
in a neighborhood of $p\in N$ we can take a local section of the
reduced frame bundle $\Redframe(N)$ such that
\begin{equation*}
    \varphi = r\,d\phi + g_{i}dy^{i}\,,
\end{equation*}
where $r$ is the radius of the circle. From this we see that
\begin{equation*}
    d\varphi = \frac{1}{r}\,dr \wedge \varphi + (\text{stuff}) dy^{0} 
    \wedge dy^{1}\,.
\end{equation*}
Comparing with \eqref{eq:SO2-dvarphi} we see that
\begin{equation}
    K_{i}\theta^{i} = \frac{dr}{r}\,.
    \label{eq:K-dr}
\end{equation}
in agreement with \eqref{eq:K-curl-0}.  Expression \eqref{eq:K-dr}
tells you that $r: \Redframe(N) \to \bbR_{+}$ is really the pullback
to the reduced frame bundle of a function $r_{M}:M \to \bbR_{+}$.  The
first and second derivatives of $r$ are defined by
\begin{equation}
    \begin{split}
    dr &= r_{i}\theta^{i}\,, \\
    dr_{i} & = -\pi_{ij}r_{j} + r_{i;j}\theta^{j}\,,
    \end{split}
    \label{eq:def-dr}
\end{equation}
where $r_{i;j}= r_{j;i}$.

The Ricci tensor given by
\begin{equation}
    \begin{split}
	R^{N}_{\varphi i} & = -\epsilon_{i}{}^{j} \left( A_{;j} + 2 
	    A\, \frac{r_{j}}{r} \right)\,, \\
	R^{N}_{ij} & = - \frac{r_{i;j}}{r} - \left( k^{M} + A^{2}
	\right) \eta_{ij}\,,\\
	R^{N}_{\varphi\varphi} & = -\frac{r^{i}{}_{;i}}{r} -2A^{2}\,.
    \end{split}
    \label{eq:Sch-Ricc}
\end{equation}

Armed with this information we apply the submersion geometry to
see what extra properties we can obtain.  The first observation is
that $R^{N}_{\varphi i}=0$ from which we learn that
\begin{equation*}
    A_{;j} + 2 A\, \frac{r_{j}}{r} =0.
\end{equation*}
This equation is trivial to solve 
\begin{equation}
    A = \frac{a}{r^{2}}\,,
    \label{eq:A-soln}
\end{equation}
where $a \in\bbR$ is a constant a constant of integration.

Next we observe that the remaining Einstein equations become
\begin{align}
    0 & = - \frac{r_{i;j}}{r} - \left( k^{M} -2\, \frac{a^{2}}{r^{4}}
    +2\Lambda\right) \eta_{ij}\,, 
    \label{eq:e-1}\\
    0 & = -\frac{r^{i}{}_{;i}}{r} -2\, \frac{a^{2}}{r^{4}} -2\Lambda\,.
    \label{eq:e-2}
\end{align}
Taking the trace of the first equation above we see that
\begin{equation}
    0 = - \frac{r^{i}{}_{;i}}{r} - 2\left( k^{M} -2\, \frac{a^{2}}{r^{4}}
    +2\Lambda \right),
    \label{eq:e-1-tr}
\end{equation}
From \eqref{eq:e-2} and \eqref{eq:e-1-tr} we learn that
\begin{equation}
    k^{M} = -\Lambda + 3\, \frac{a^{2}}{r^{4}}\,.
    \label{eq:e-4}
\end{equation}
Note that $M$ has a curvature singularity as $r \to 0$.
Note that \eqref{eq:e-1} becomes
\begin{equation}
     \frac{r_{i;j}}{r}=  -\left( \Lambda + \frac{a^{2}}{r^{4}}
    \right) \eta_{ij}
    \label{eq:e-3}
\end{equation}

The Cartan structural equations for the reduced frame bundle
$\Redframe(N)$ are
\begin{align}
    d\theta^{0} & = +\pi \wedge \theta^{1}\,,
    \label{eq:SO2-dtheta0-1}  \\
    d\theta^{1} & = +\pi \wedge \theta^{0}\,,
    \label{eq:SO2-dtheta1-1} \\
    d\pi & = d\pi_{01} =\left(-\Lambda + \frac{3a^{2}}{r^{4}}\right)
    \theta^{0} \wedge \theta^{1}\,, 
    \label{eq:SO2-dpi-1} \\
    d\varphi  &= +\frac{1}{r}\; dr \wedge \varphi - 
	\frac{2a}{r^{2}}\; \theta^{0}\wedge\theta^{1}\,.
    \label{eq:SO2-dvarphi-1}
\end{align}
These four equations have an interesting structure.  The first three
equations are a closed system of equations and define the Lorentz
frame bundle $\Redframe(M)$ of the base manifold $M$ with the
Levi-Civita connection.  We see that this manifold has a potential
curvature singularity when $r=0$.  The Frobenius theorem tell us that
we have a foliation defined by the exterior differential system
$\theta^{0}=\theta^{1}=\pi=0$.  On the one dimensional leaves we have
$d(\varphi/r)=0$ and therefore we have that $\varphi = r\, d\phi$ when
restricted to the leaf for some angular coordinate $\phi$.  In plain
language, we construct $\Redframe(M)$ using the first three structural
equations.  Subsequently we use the fourth equation to construct the
full reduced frame bundle $\Redframe(N)$.  Studying the geometrical
properties of $M$ will give us a lot of information about $N$.

\subsection{Properties of the radius function}
\label{sec:radius-function}

Next we derive a differential equation satisfied by $\nu=r^{i}r_{i} =
\lVert dr \rVert^{2}_{M}$. In the study of the Schwarzschild solution 
we saw that the critical points of $r$ played a central role.
\begin{equation*}
    \begin{split}
	d \nu &= 2 r^{i}r_{i;j}\theta^{j}\,, \\
	 &= -2\left(\Lambda r + \frac{a^{2}}{r^{3}} \right) dr\,.
    \end{split}
\end{equation*}
The solution to this differential equation is elementary and given by
\begin{equation}
    \nu = \lVert dr \rVert^{2}_{M} = -b -\Lambda r^{2} + 
    \frac{a^{2}}{r^{2}}\,,
    \label{eq:S2-sol}
\end{equation}
where $b \in \bbR$ is a constant of integration.  

We note that generically there is no asymptotic Minkowski region as
$r\to\infty$.  In such a region we should have $\lVert dr
\rVert_{M}^{2} \to 1$ and this requires $b=-1$ and $\Lambda=0$.  The
manifold $N$ is flat but the horizontal spaces of the
submersion $N \to M$ are not integrable if $a\neq 0$.  Note that $r$
does not have critical points if $b=-1$, $\Lambda=0$.

At a critical point of $r$ we have that $dr=0$ and therefore $\lVert
dr \rVert_{M}^{2} =0$.  Because the metric has lorentzian signature
the converse is not true: $\lVert dr \rVert_{M}^{2} =0$ does not imply
$dr=0$.  From \eqref{eq:S2-sol} we see that $dr$ is a null $1$-form at
\begin{equation}
    \rho_{\pm}^{2} = \frac{-b \mp \sqrt{b^{2} + 4 \Lambda 
    a^{2}}}{2\Lambda}\,.
    \label{eq:rpm}
\end{equation}
Physics and mathematics requires that the roots satisfy $\rho_{\pm}^{2} \ge 0$. 
This leads to various cases:
\begin{enumerate}
    \item There are no acceptable roots if the discriminant $b^{2} +
    4\Lambda a^{2} <0$.

    \item  There may be acceptable roots if the discriminant $b^{2} + 
    4\Lambda a^{2} \ge 0$.
\end{enumerate}

In the BTZ black hole we have a negative cosmological constant
$\Lambda = -1/\ell^{2}$.  Comparing with BTZ we see that $b=M$, the
mass of the black hole, and $2a=J$, the angular momentum.  We see that
$\rho_{\pm}^{2} = \tfrac{1}{2} \ell^{2} \left( M \pm \sqrt{M^{2}
-J^{2}/\ell^{2}}\right)$ and for the existence of critical point of
$r$ we require $\lvert J \rvert \le M\ell$. We also note using 
\eqref{eq:K-dr} that the extrinsic curvature has norm
\begin{equation*}
    \lVert K \rVert^{2}_{M} = \frac{\lVert dr \rVert^{2}_{M}}{r^{2}} =
    -\frac{M}{r^{2}} + \frac{1}{\ell^{2}} + \frac{J^{2}}{4r^{2}}\,.
\end{equation*}
that is well defined as $r\to \infty$. Note that $K$ is null if and 
only if $dr$ is null.
$r$ we require $\lvert J \rvert \le M\ell$.  The so called outer and
inner horizons of the BTZ black hole are located at $r=\rho_{+}$ and
$r=\rho_{-}$ respectively.  The standard notation is to use $r_{\pm}$
for $\rho_{\pm}$ but in this article we follow the notation of
\cite{Alvarez:2007xs} where $r_{\pm}$ are used for the derivatives of
$r$ in the null directions.

\section{BTZ Killing Vectors}
\label{sec:BTZ-Killing}

We point out that automatically there is an extra killing
vector besides the one that generates the $\SOrth(2)$ action.  
Consider a general vector field
\begin{equation*}
    X = X^{i} \vece_{i} + X^{\varphi} \vece_{\varphi} + \half X^{ij}
    \vece_{ij}
\end{equation*}
then we have
\begin{align}
    \Lieder_{X}\varphi &= dX^{\varphi} +
    \frac{X^{i}r_{i}}{r} \, \varphi - X^{\varphi}\, \frac{dr}{r} 
    \nonumber \\
    & \quad -\frac{2a}{r^{2}}\, \left(X^{0}\theta^{1} - 
    X^{1}\theta^{0}\right)\,, 
    \label{eq:Lie-met-1} \\
    &= r\, d\left(X^{\varphi}/r\right) +
    \frac{X^{i}r_{i}}{r} \, \varphi 
    \nonumber \\
    &\quad -\frac{2a}{r^{2}}\, \left(X^{0}\theta^{1} - 
    X^{1}\theta^{0}\right)\,, 
    \label{eq:Lie-met-1a}  \\
    \Lieder_{X} \theta^{i} &= -X_{ij}\theta^{j} + 
    DX^{i}\,.
    \label{eq:Lie-met-2}
\end{align}
As a passing remark we note that the $\SOrth(2)$ Killing vector is
easily seen by inspection to be
\begin{equation}
    X = r \vece_{\varphi}.
    \label{eq:so2-killing}
\end{equation}

First we look for solutions to the Killing equations coming from
\eqref{eq:Lie-met-2}.  We note that $DX_{i} = X_{i;j}\theta^{j} +
X_{i;a} \theta^{a}$.  The Killing conditions require $X_{i;a}=0$,
\emph{i.e.}, $X^{i}$ is intrinsically associated with the base $M$.
Note that the $\SOrth(2)$ Killing vector $r\vece_{\varphi}$ on $N$
projects to zero on $M$.  If we take one of those $X^{i}$ Killing
vectors related to the Lie group $\Orthframe(M)$ and try to lift to
$\Redframe(N)$ by plugging into the Killing equation associated with
\eqref{eq:Lie-met-1} then we see that the $X^{i}$ have to be chosen to
have some type of relationship with the function $r$.  To understand
this best look at the Killing equations coming from
\eqref{eq:Lie-met-1a}
\begin{equation*}
    d\left(X^{\varphi}/r\right) +
        \frac{X^{i}r_{i}}{r^{2}} \, \varphi 
        -\frac{2a}{r^{3}}\, \left(X^{0}\theta^{1} - 
        X^{1}\theta^{0}\right) =0\,.
\end{equation*}
Look at the integrability conditions by taking the exterior derivative
\begin{equation*}
    d \left[
	\frac{X^{i}r_{i}}{r^{2}} \, \varphi 
	-\frac{2a}{r^{3}}\, \left(X^{0}\theta^{1} - 
	X^{1}\theta^{0}\right) \right] =0\,.
\end{equation*}
This integrability equation is independent of $X^{\varphi}$.  This
equation gives \emph{algebraic} relations between the $X^{i}$, $X_{01}$ and
the function $r$ and its derivatives.  This means that a generic
Killing vector on $M$, that knows nothing about $r$, will not lift to
a Killing vector on $N$.  We have to look for Killing vectors on $M$
that are compatible with the $r$ dependence that appears in
$\Redframe(N)$.  Let's build this into an ansatz for the $X^{i}$.  We
note that generically, \emph{i.e.}, when $\lVert dr \rVert_{M} \neq
0$, that $dr$ and $*dr$ are linearly independent.  The Cartan
structural equations tell us that $r:\Redframe(N) \to \bbR$ is
essentially the only object we have to play with.  It is best to work
in a light cone frame.  We will choose $X_{+} = r_{+} F(r)$ and $X_{-}
= r_{-} G(r)$.  We have to solve $X_{i;j} + X_{j;i} = 0$.  For the
moment we do not need $X^{\varphi}$.  We note that \eqref{eq:e-3}
tells us that $r_{+;+}=0$.  We have Killing's equation
\begin{equation*}
    0 = X_{+;+} = r_{+;+} F(r) + r_{+}r_{+}F'(r)= r_{+}^{2} F'(r).
\end{equation*}
We immediately learn that $F$ is constant. Likewise from the 
$X_{-;-}=0$ equation we learn that $G$ is constant. Finally we 
observe that $0=X_{+;-} + X_{-;+}$ tells us that $F=-G$. Thus we 
conclude that $X_{\pm} = \pm F r_{\pm}$. We choose the normalization
\begin{equation}
    X^{i} = - \lambda \epsilon^{ij} r_{j}\quad \text{where }
    \lambda\in\bbR\,.
    \label{eq:Killing-extra}
\end{equation}
The vectors $r^{i}$ and $X^{i}$ are Minkowski orthogonal, $r_{i} X^{i}
=0$, and that
\begin{equation}
    \left\lVert X \right\rVert^{2}_{M} = - \lambda^{2} \left\lVert
    \bnabla r \right\rVert^{2}_{M}\,.
    \label{eq:S-X-norm}
\end{equation}
Note that if $\bnabla r$ is
spacelike then $X$ is timelike and vice-versa.  If $\bnabla r$ is lightlike
then $X$ is also lightlike and vice-versa. 

Next we plug $X^{i}$ into the Killing equation coming from 
\eqref{eq:Lie-met-1a} to obtain
\begin{equation*}
    d\left(X^{\varphi}/r\right)  
	-\frac{2a}{r^{3}}\, \lambda\, dr =0\,.
\end{equation*}
This equation is trivial to integrate yielding
\begin{equation}
    X^{\varphi} = \mu r - \lambda \, \frac{a}{r}\,,
    \label{eq:X-phi-soln}
\end{equation}
where $\mu \in \bbR$ is a constant of integration. Thus we get a  two 
parameter family of Killing vector fields on $N$. Associated with the 
$\SOrth(2)$ action we have
\begin{equation*}
    X_{\SOrth(2)} = r \vece_{\varphi}\,.
\end{equation*}
The other Killing vector is given on $\Redframe(N)$ by
\begin{equation}
    T = -\epsilon^{ij}r_{j}\vece_{i} - \frac{a}{r}\,\vece_{\varphi}
    - \left(\Lambda r + \frac{a^{2}}{r^{3}}\right) \vece_{01}\,.
    \label{eq:Killing-time}
\end{equation}
We denote this vector by $T$ to remind the reader that in the 
``ordinary'' region it is the timelike Killing vector associated with ``time 
translations''. We note the norms of this vector when projected to 
$M$ and $N$ are respectively given by
\begin{align}
    \lVert T \rVert^{2}_{M} & = -\left( -b -\Lambda r^{2} + 
    \frac{a^{2}}{r^{2}} \right) = -\lVert dr \rVert_{M}^{2},
    \label{eq:T2-M}  \\
    \lVert T \rVert^{2}_{N} &  = -\left( -b -\Lambda r^{2} + 
    \frac{a^{2}}{r^{2}} \right) + \left(\frac{a}{r}\right)^{2}
    \nonumber \\
    &\quad  =
    b + \Lambda r^{2}\,.
    \label{eq:T2-N}
\end{align}

It is well known that if you have a timelike Killing vector then the 
redshift between at emitter E and an observer O is given by
\begin{equation}
    \frac{\omega_{O}}{\omega_{E}} = \sqrt{\frac{\lVert T_{E}
    \rVert^{2}_{N}}{\lVert T_{O} \rVert^{2}_{N}}}\,.
    \label{eq:redshift}
\end{equation}
From this we see that an observer sees an infinite redshift if
the photon is emitted at a location where the Killing vector becomes
null $\lVert T_{E} \rVert^{2}_{N} =0$.  This is not the necessarily on
the event horizon. For the BTZ black hole the infinite redshift 
surface is located at $r$ given by $\rho_{\infty}^{2} = M\ell^{2} = 
\rho_{+}^{2} + \rho_{-}^{2}$.

\section{Geodesics in M}
\label{sec:geodesics-1}

We already discussed that there is very close relationship between
$\Redframe(M)$ and $\Redframe(N)$, namely there is a fibration
$\Redframe(N) \to \Redframe(M)$.  The study of the horizontal curves
in $\Redframe(N)$ will allow us to probe the geometry of the base
manifold and give us information about geometry of the BTZ spacetime.
We study horizontal curves on $\Redframe(N)$ that will have the form
$u^{+} \vece_{+} + u^{-} \vece_{-} + u^{\varphi}\vece_{\varphi}$ where
$u^{+}$, $u^{-}$ and $u^{\varphi}$ are constant.  We restrict to the
special case $u^{\varphi}=0$.  The curve in this case may be viewed as
a horizontal curve on the lorentzian frame bundle $\Redframe(M)$ that
projects down to a geodesic on $M$.  We work out some properties of
the geodesics on the base $M$ by using the exponential map 
\emph{\`{a} la} Cartan~\cite{Cartan:riemann} that is also described in
\cite{Alvarez:2007xs}.

We introduce a null basis for the canonical $1$-forms on the Lorentz
frame bundle of $M$ by defining $\theta^{\pm} = \theta^{0} \pm
\theta^{1}$.  We do not need all the details of the exponential map.
All we need is the behavior of the radius function and its derivatives
along the geodesic.  Let $\lambda$ be an affine parameter along the
geodesic.  From this it follows that
\begin{equation}
    dr/d\lambda = r_{+}u^{+} + r_{-}u^{-}\,.
    \label{eq:dr-dlambda} 
\end{equation}
We also need  $r_{+;+}=r_{-;-}=0$ by \eqref{eq:e-3} to show that
\begin{equation}
    \begin{split}
	\frac{d r_{+}}{d \lambda} &= \frac{1}{2}\left( \Lambda r +
	\frac{a^{2}}{r^{3}} \right) u^{-}\,, \\
	\frac{d r_{-}}{d \lambda} &=
	\frac{1}{2}\left( \Lambda r + \frac{a^{2}}{r^{3}} \right) u^{+}\,.
    \end{split}
    \label{eq:dr-evolution}
\end{equation}
Consequently we see that
\begin{equation}
    \frac{d^{2}r}{d\lambda^{2}} = -\left( \Lambda r +
    \frac{a^{2}}{r^{3}} \right) \left\lVert u \right\rVert_{M}^{2}.
    \label{eq:r-ODE}
\end{equation}

The discussion here is taken almost verbatim from 
\cite{Alvarez:2007xs}.
The case of a null radial geodesic is particularly simple because
$d^{2}r/d\lambda^{2}=0$.  If the horizontal lift of the null geodesic
begins at a point $p\in \Redframe(M)$ with $r(p) = r_{p}$ and $dr(p)
= r_{i}(p) \theta^{i}(p)$ then the evolution of $r$ along the lift is
\begin{equation}
    r(\lambda) = r_{p} + \lambda\left( r_{+}(p) u^{+} + r_{-}(p) u^{-} 
    \right) .
    \label{eq:soln-ODE-r}
\end{equation}
There are four cases of null geodesics to analyze corresponding to
\begin{equation*}
    (u^{+},u^{-}) \in \left\{ (+1,0),(0,+1),(-1,0),(0,-1) \right\}\,.
\end{equation*}
The latter two cases may be considered with the first two by allowing
$\lambda$ to be negative.  In the first case we have that $r(\lambda)
= r_{p} + \lambda r_{+}(p)$, and in the second case we have
$r(\lambda) = r_{p} + \lambda r_{-}(p)$.  Choose a Lorentz frame
$p\in\Orthframe(M)$, if $r_{+}(p)>0$ then $r_{+}(p') >0$ for all $p'$
in the same fiber because the action of the $(1+1)$ dimensional
Lorentz group translates to an action $r_{\pm} \to e^{\pm\eta}r_{\pm}$
where $\eta$ is the rapidity.  This means that we can define the
following four open subsets of $M$:
\begin{equation}
    \begin{split}
        U_{\text{I}} &= \{ q \in M \;|\; r_{+}(p)>0, r_{-}(p) <0\}\,, \\
	U_{\text{II}} &= \{ q \in M \;|\; r_{+}(p)<0, r_{-}(p) <0\}\,, \\
	U_{\text{III}} &= \{ q \in M \;|\; r_{+}(p)>0, r_{-}(p) >0\}\,, \\
	U_{\text{IV}} &= \{ q \in M \;|\; r_{+}(p)<0, r_{-}(p) >0\}\,.
    \end{split}
    \label{eq:regions}
\end{equation}
In the above $p \in \Orthframe(M)$ is any Lorentz orthonormal frame at
$q\in M$.

We assume our space-time manifold $N$ has a ``normal region'' where a
light ray can go radially inward with initial condition
$(u^{+},u^{-})=(0,1)$ or radially outward with initial condition
$(u^{+},u^{-})=(1,0)$ and ``contains'' $r=\infty$ in a way we will
clarify later.  In such a region we can choose a $p\in \Orthframe(M)$
with the property that $r_{+}(p)>0$ and $r_{-}(p) <0$ and thus we
conclude that $U_{\text{I}} \neq \emptyset$ and that the ``normal
region'' lies in $U_{\text{I}}$.  According to \eqref{eq:soln-ODE-r},
an inward future directed radial null geodesic will have $r(\lambda) =
r_{p} + \lambda r_{-}(p)$.  The important observation is that for
finite positive affine parameter the light ray will hit $r=0$.  This
last observation says that our space may have a singularity because
the Cartan structural equations have a singularity at $r=0$.  We will
not address the question of whether this is a real or a removable
singularity because this is discussed in detail in
\cite{Banados:1992gq}.

\section{BTZ Geometry without Coordinates}
\label{sec:BTZ-geom}

The key to understanding the geometry of the BTZ solution is
to understand the level sets of the radius function $r: M \to
\bbR_{+}$.  For all practical purposes, both physical and
mathematical, we can take $M$ to be simply connected.  Here we 
construct the global structure of the BTZ spacetime.

\begin{figure}[tbp]
    \raggedright
    \includegraphics[width=0.4\textwidth]{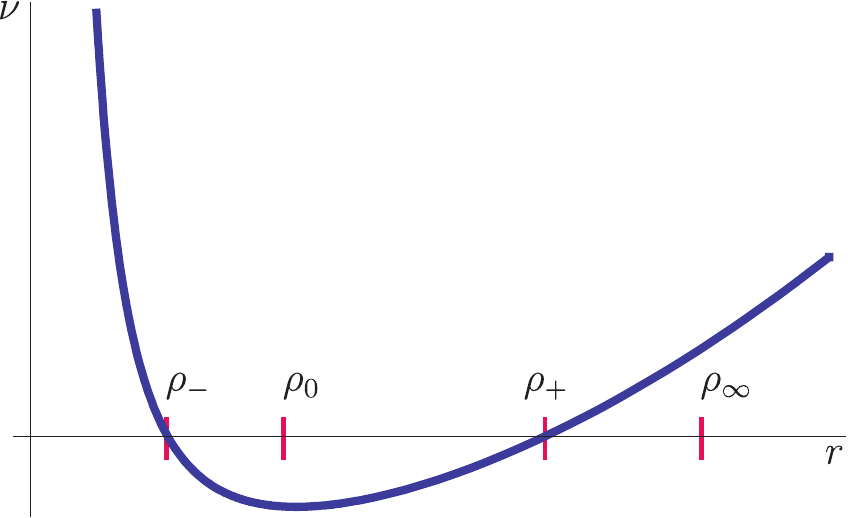}
    \caption[x]{\small Behavior of the function $\nu = \lVert dr
    \rVert^{2}_{M}$ for $0 <r<\infty$ where $0< 2a <bl$.  Note
    that $\rho_{-} < \rho_{0} < \rho_{+} < \rho_{\infty}$.  Here
    $\rho_{0}^{2}=a\ell=J/2\ell$ is minimum of $\nu$.  The ergocircle, the
    curve of infinite redshift, is located at $\rho_{\infty}^{2}=
    b\, \ell^{2}=M\ell^{2}$.  The infinite redshift condition is
    $\lVert T \rVert^{2}_{N}=0$.}
    \label{fig:nu}
\end{figure}

If $r:M \to \bbR_{+}$ has critical points then they must be
non-degenerate because of \eqref{eq:e-3}.  We assume that $N$ is
$\adS_{3}$ and define $\Lambda = -1/\ell^{2}$.  The reader is reminded
that BTZ showed that $b=M$ and $a=J/2$.  Rewriting we have
\begin{equation}
    \nu = \lVert dr \rVert_{M}^{2}= -\lVert T \rVert_{M}^{2} = -b +
    \frac{r^{2}}{\ell^{2}} + \frac{a^{2}}{r^{2}}\,.
    \label{eq:nu-2}
\end{equation}
The general shape of this function is shown in Figure~\ref{fig:nu}.

The $1$-form $dr$ is null at
\begin{equation}
    \rho^{2}_{\pm} = \frac{1}{2}\; b\ell^{2} \left( 1 \pm 
    \sqrt{1-\frac{4a^{2}}{b^{2}\ell^{2}}} \right).
    \label{eq:rpm-2}
\end{equation}
Note that
\begin{equation*}
    \frac{1}{\rho^{2}_{\pm}} = \frac{b}{2a^{2}}\; \left( 1 \mp
    \sqrt{1-\frac{4a^{2}}{b^{2}\ell^{2}}} \right).
\end{equation*}
We also note that
\begin{equation}
    r_{i;j} = \frac{1}{r} \left( \frac{r^{2}}{\ell^{2}} - 
    \frac{a^{2}}{r^{2}} \right) \eta_{ij}\,.
    \label{eq:r-ij-2}
\end{equation}
Thus we have
\begin{equation}
    r_{i;j}(\rho_{\pm}) = \pm\frac{b}{\rho_{\pm}}
    \sqrt{1-\frac{4a^{2}}{b^{2}\ell^{2}}}\; \eta_{ij}\,.
    \label{eq:r-ij-3}
\end{equation}

Assume the radius function $r: M \to \bbR_{+}$ has a critical point at
$p \in M$.  We know by \eqref{eq:r-ij-3} that this critical point is
non-degenerate if $4 a^{2} < b^{2}\ell^{2}$.  We note that the minimum
of $\nu$ occurs at $\rho^{2}_{0}=a\ell$ and curiously
$r_{i;j}(\rho_{0})=0$.  The extremal BTZ black hole with $b\ell=2a$
($J=M\ell$) have $\rho_{\pm}^{2} = \rho_{0}^{2}$ and
$r_{i;j}(\rho_{0}) =0$ and therefore the following discussion as
presented will break down because it strongly relies on the critical
point of $\nu$ being non-degenerate.

Morse's lemma~\cite{Milnor:morse} tells us that in a neighborhood of a
critical point $p$ we can find local coordinate $(y^{0},y^{1})$
centered at $p$ that are Minkowski orthonormal at $p$ and in that
neighborhood
\begin{align}
    r(y) &= \rho_{\pm} \pm\frac{b}{2\, \rho_{\pm}}
    \sqrt{1-\frac{4a^{2}}{b^{2}\ell^{2}}} 
    \nonumber \\
    &\quad \times \left[-(y^{0})^{2} +
    (y^{1})^{2} \right]\,.
    \label{eq:r-Morse}
\end{align}

\begin{figure}[tbp]
    \raggedright
    \includegraphics[width=0.42\textwidth]{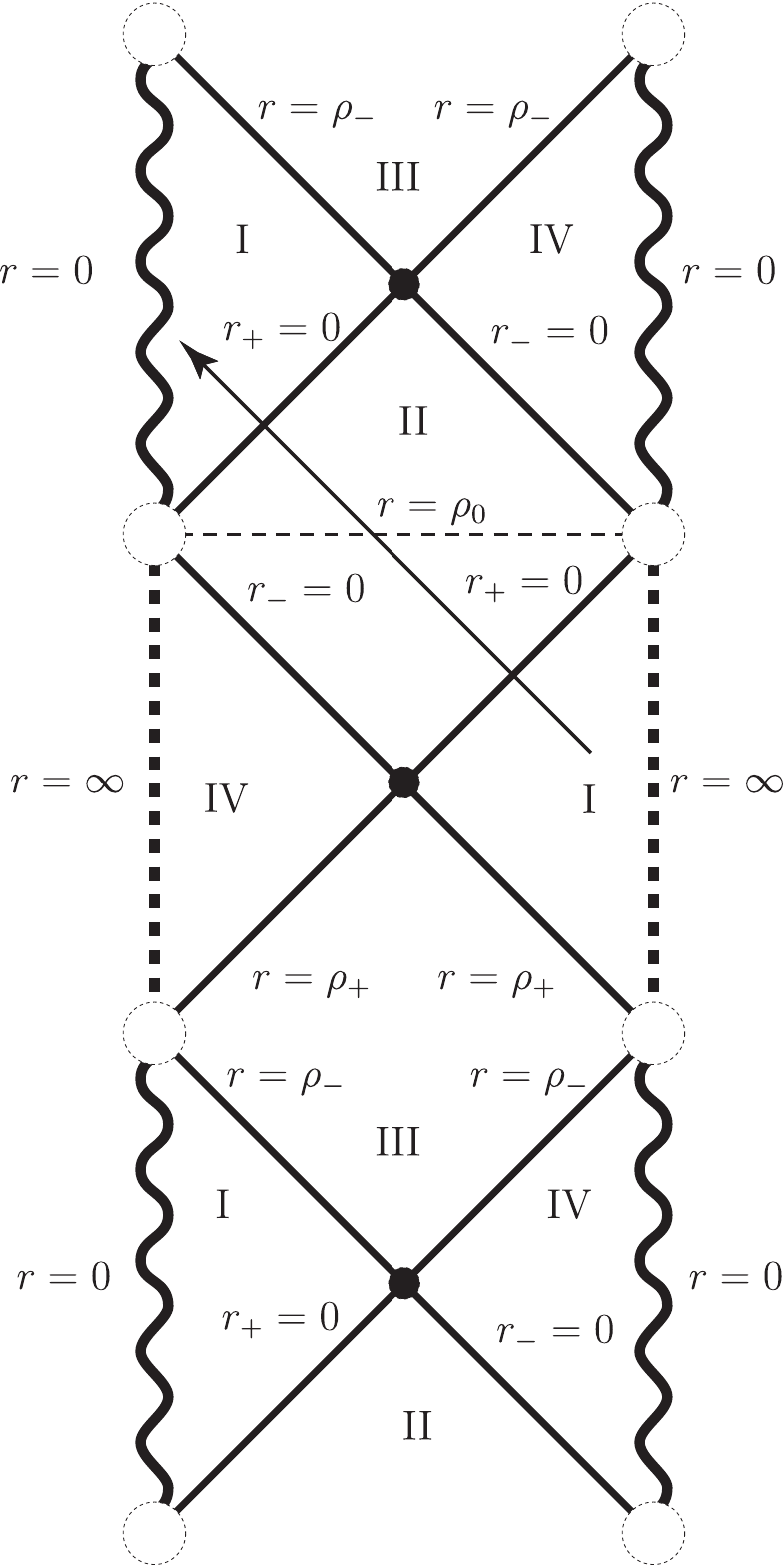}
    \caption[xy]{\small This is a very busy Penrose-Carter diagram. 
    The various curves are level sets of the function $r$. Curves 
    where $r_{\pm}=0$ are also indicated. The critical points are the 
    black circles. The arrow denotes an incoming null geodesics 
    starting in a region of type I ``containing'' $r=\infty$. Note 
    that there are two kinds of regions of type I and IV; those 
    containing $r=\infty$ and those containing $r=0$.}
    \label{fig:Penrose}
\end{figure}

Figure~\ref{fig:Penrose} is the Carter-Penrose diagram for the BTZ
black hole.
Here we describe how to construct this diagram.  By hypothesis we
start an excursion at the tail of the arrow on an inward bound future
directed null geodesic that begins in a normal region if type I
containing $r=\infty$.  This geodesic has $(u^{+},u^{-}) = (0,1)$ and
moves in the NW direction.  According to \eqref{eq:soln-ODE-r} the
radius $r$ decreases along this geodesic.  From
\eqref{eq:dr-evolution} we see that $r_{-}$ is constant along this
curve and $r_{+}$ is decreasing.  When we get to $r=\rho_{+}$ we find
that $r_{+}=0$ and we stop.  This is where we cross the level set for
$r_{+}=0$ as indicated in the figure.  Note that $\lVert dr
\rVert_{M}^{2} = -4r_{+}r_{-}$ and this one way of concluding that $r
= \rho_{+}$.  Now we make a side excursion to locate a critical point
of $r$.  Since $r_{-}$ is constant along our geodesic we conclude that
$r_{-}<0$ where we are.  Chose a null geodesic with $(u^{+},u^{-}) =
(-1,0)$ and start moving (SW direction).  Note that $r$ is constant
and $r_{+}=0$ along this geodesic.  Equation \eqref{eq:dr-evolution}
tells us that $dr_{-}/d\lambda>0$ along this SW directed geodesic.  We
stop when $r_{-}=0$ and we have found our nondegenerate critical
point.  Note that if our side excursion had chosen $(u^{+},u^{-}) =
(1,0)$ (NE direction) then $dr_{-}/d\lambda <0$ and we will not hit a
critical point.  Since the critical point is nondegenerate we know the
there must be a $r_{-}=0$ level set emanating from it (see
\eqref{eq:r-Morse}).  We now go back to where we began the side
excursion and continue along the the NW arrow.  We note that $r_{-}<0$
remains constant along this trajectory.  The key observation is that
$r_{+}$ begins to decrease and goes negative while we are near
$r=\rho_{+}$ but equation \eqref{eq:dr-evolution} tells us that once
we cross $r=\rho_{0}$ the sign of the right hand side of the
$dr_{+}/d\lambda$ equation changes sign.  This means that $r_{+} <0$
begins to increase and reaches $r_{+}=0$ when $r=\rho_{-}$.  Now we
are ready for our second side excursion.  You can verify that if you
go along the null geodesic in the NE direction then $r_{-}$ increases
and you will eventually get to $r_{-}=0$ and you have found another
nondegenerate critical point.  There is no critical point in the SW
direction because $r_{-}$ would be decreasing.  Now we go back to the
original geodesic and continue into a region of type I that contains
$r=0$.  In finite affine parameter we hit $r=0$.  Note that it is
possible to escape to infinity by stopping and getting onto a null
geodesic in the NE direction and getting away.  Notice that you will
eventually wind up in a different type I region containing
$r=\infty$ that is not the original one. 

Using this procedure and going forwards and backwards in time you can
construct the Penrose diagram for the BTZ spacetime.

\section*{Acknowledgments}
I would like to thank  Laurent Baulieu, Eliezer Rabinovici, Jan de
Boer, Michael Douglas, Pierre Vanhove and Paul Windey for giving me the
opportunity to present these lectures. I would also like to thank 
Elena Gianolio for her assistance.

\appendix

\section{adS$_{3}$ Basics}
\label{sec:adS3-Basics}

Three dimensional anti-deSitter space is the coset manifold $\adS_{3}
= \SOrth(2,2)/\SOrth(1,2)$.  This can be regarded as the
``hyperboloid'' surface
\begin{equation}
    -u^{2}-v^{2} + x^{2} + y^{2} = -\ell^{2}
    \label{eq:hyperboloid}
\end{equation}
in $\bbR^{2,2}$.  Note that this surface contains timelike circles.
In fact if we define $t = \sqrt{u^{2}+v^{2}}$ and $r = \sqrt{x^{2}
+y^{2}}$ then $t^{2}-r^{2} = \ell^{2}$.  Note that we can choose $t =
\ell \cosh\eta$ and $r = \ell \sinh \eta$ where $\eta\ge 0$.  
Roughly, $t$ is the radius of the timelike circle and $r$ is the 
radius of the spacelike circle. This
leads to a simple parametric description of the surface
\begin{align*}
    u & = \ell \cosh\eta \cos\phi\,,  \\
    v & = \ell\cosh\eta \sin\phi\,,  \\
    x & = \ell\sinh\eta \cos\theta \,, \\
    y & = \ell\sinh \eta \sin\theta\,,
\end{align*}
where $\eta \in [0,\infty)$, $\phi \in [0,2\pi]$ and $\theta \in 
[0,2\pi]$. Note that the timelike circles associated with $\phi$ are not 
contractible because the radius is bounded from below by $\ell$. The 
circles associated with $\theta$ are contractible. This means that the 
topology of $\adS_{3}$ is that of $S^{1} \times \bbR^{2}$.

Technically, what is usually called $\adS_{3}$ is the universal cover
of the above obtained by unwrapping the circle parametrized by $\phi$.

\relax


\providecommand{\href}[2]{#2}\begingroup\raggedright\endgroup

\end{document}